\documentclass[fleqn,usenatbib]{mnras}
\usepackage{graphics}
\usepackage{graphicx}
\usepackage{color}
\usepackage{aas_macros}
\usepackage{natbib}
\usepackage{amsmath}
\usepackage{comment}
\usepackage{soul} 

\newcommand{\avg}[1]{\left\langle{#1}\right\rangle}
\newcommand{\unit}[1]{\textnormal{#1}}
\newcommand{\Hunit}{\ensuremath{\frac{\unit{km}/\unit{s}}{\unit{Mpc}}}}

\title{Concordance cosmology without dark energy}

\author[G. R\'acz et al.]{G\'abor R\'acz$^1$\thanks{E-mail: ragraat@caesar.elte.hu},  L\'aszl\'o Dobos$^1$,  R\'obert Beck$^1$, Istv\'an Szapudi$^2$, Istv\'an Csabai$^1$
\\
$^1$Department of Physics of Complex Systems, E\"{o}tv\"{o}s Lor\'and University, Pf. 32, H-1518 Budapest, Hungary\\
$^2$Institute for Astronomy, University of Hawaii, 2680 Woodlawn Drive, Honolulu, HI, 96822
}

\date{Accepted XXX. Received YYY; in original form ZZZ}

\pubyear{2017}

\begin{document}
\label{firstpage}
\pagerange{\pageref{firstpage}--\pageref{lastpage}}
\maketitle

\begin{abstract}

According to the separate universe conjecture, spherically symmetric sub-regions in an isotropic universe behave like mini-universes with their own cosmological parameters. This is an excellent approximation in both Newtonian and general relativistic theories. We estimate local expansion rates for a large number of such regions, and use a scale parameter calculated from the volume-averaged increments of local scale parameters at each time step in an otherwise standard cosmological $N$-body simulation. The particle mass, corresponding to a coarse graining scale, is an adjustable parameter. This mean field approximation neglects tidal forces and boundary effects, but it is the first step towards a non-perturbative statistical estimation of the effect of non-linear evolution of structure on the expansion rate. Using our algorithm, a simulation with an initial $\Omega_m=1$ Einstein--de~Sitter setting closely tracks the expansion and structure growth history of the $\Lambda$CDM cosmology. Due to small but characteristic differences, our model can be distinguished from the $\Lambda$CDM model by future precision observations. Moreover, our model can resolve the emerging tension between local Hubble constant measurements and the \textit{Planck} best-fitting cosmology. Further improvements to the simulation are necessary to investigate light propagation and confirm full consistency with cosmic microwave background observations.

\end{abstract}


\maketitle

\section{Introduction}

Gravitation being the only effective force on the largest scales, cosmological evolution is governed by general relativity (GR). To zeroth order, the homogeneous and isotropic Friedmann--Lema\^{\i}tre--Robertson--Walker (FLRW) solutions of Einstein's equations drive the expansion and growth history of the Universe. The concordance $\Lambda$CDM model \citep[e.g.,][]{Planck2015} posits an unknown form of energy with negative pressure and an energy density about $10^{123}$ off from theoretical expectations. The $\Lambda$CDM paradigm reproduces most observations, although, to this day, no plausible candidate for dark energy has emerged and some tensions remain \citep[see][for a recent comprehensive review]{Buchert_2016IJMPD..2530007B}. Most notably, the latest local measurements \citep{Riess_hubbleParameterTension_2016} of the Hubble constant are up to $3.4\sigma$ high compared to the value derived from \textit{Planck} observations \citep{Planck2015} of the cosmic microwave background.

The ubiquitous presence of clusters, filaments, and voids in the cosmic web manifestly violate the assumed homogeneity of $\Lambda$CDM. Given the non-linear nature of Einstein's equations, it has been known for a while that local inhomogeneities influence the overall expansion rate, whereas the magnitude of such backreaction effect is debated. In particular, \citet{2014CQGra..31w4003G,2016CQGra..33l5027G} argued that the effect of inhomogeneities on the expansion of the Universe is irrelevant, while \citet{2015CQGra..32u5021B} disputed the general applicability of the former proof. More recently, \citet{PhysRevLett.116.251301} used numerical relativity to show the existence of a departure from FLRW behaviour due to inhomogeneities, beyond what is expected from linear perturbation theory. Nevertheless, the spectacular successes of the homogeneous concordance model suggest that any effect of the inhomogeneities on the expansion rate should be weak, {\em unless} it mimics the $\Lambda$CDM expansion and growth history to a degree allowed by state of the art observations. In this spirit, we present a statistical non-perturbative algorithm, a simple modification to standard $N$-body simulations, that provides a viable alternative to dark energy while it can simultaneously resolve the Hubble constant puzzle.

In the late-time non-linear evolution of the Universe, coarse graining and averaging are both problematic \citep[see][and references therein]{Wiltshire2007a, WiltshireCQG}. The complexity of Einstein's equations prevents direct numerical modelling of backreaction, which ideally would require a general relativistic simulation of space-time seeded with small initial fluctuations. Such an ideal simulation would contain a hierarchy of coarse graining scales describing the space-time of stars, galaxies, galaxy clusters, intra-cluster medium and dark matter, etc., and their metric would be stitched together in a careful manner \citep[see][for a discussion of the hierarchy of coarse graining scales]{Wiltshire2014}. As fluctuations grow due to non-linear gravitational amplification, space-time itself becomes complex, and even the concept of averaging becomes non-trivial.

Despite the difficulties, \citet{Buchert2000, Buchert2001} and others \citep{2012ARNPS..62...57B} realized that backreaction can be understood in a statistical fashion through the spatial averaging of Einstein's equations on a hypersurface, leading to the Buchert equations.
Second-order perturbative solutions to the equations are given by \cite{Kolb2005} and \cite{Rasanen2010}, while \cite{Wiltshire2007a, Wiltshire2007b, Wiltshire2009} presents an analytical solution to a two-scale (voids and walls) inhomogeneous universe.

We propose a non-perturbative, multi-scale statistical approach to study strong backreaction based on the general relativistic separate universe conjecture, which states that a spherically symmetric region in an isotropic universe behaves like a mini-universe with its own energy density $\Omega = 1 + \delta$ \citep{Weinberg2008}. The conjecture was proven by \cite{Dai2015} for compensated top hat over- and underdensities. In the quasi Newtonian framework, the separate universe conjecture is widely used in successful spherical collapse models \citep{Bernardeau1994, Mohayaee2006, 2016MNRAS.455L..11N}. We build on this conjecture to estimate the expansion rate as the volume average of local expansion rates, avoiding the calculation of any geometric quantities such as the average curvature of the universe. Our algorithm neglects tidal forces, but the scales over which averages are calculated are solidly grounded in Newtonian physics, simplifying the interpretation of our results. We show that under our algorithm, the non-Gaussian distribution of matter arising from the non-linearities of the cosmological fluid equations causes the expansion rate to decrease at a slower rate than normally calculated from the global Friedmann equations, thereby mimicking the effect of dark energy.

In our scheme the coarse graining scale is an adjustable, phenomenological parameter corresponding to the best ``particle size'' to use when modelling the evolution of the Universe. When the coarse graining scale approaches the scale of homogeneity, our model, obviously, shows no effect: in this limit it is equivalent to the global Friedmann equations.  Approaching very small scales, the assumptions of the model progressively break down due to the increasing anisotropy around, and inhomogeneity inside, the spherically symmetric regions. Somewhere between the extremes, there is an optimal scale that we expect to be around the size of virialized structures detached from the Hubble flow, therefore on the order of $10^9-10^{13} M_{\odot}$. The coarse graining scale is a semi-nuisance parameter to be fit, analogous to halo model parameters. While in this work we use a single, redshift-independent, comoving coarse graining scale, in principle, the optimal scale could depend on the state of the Universe and its constituents and thus, on redshift.

\section{Inhomogeneous model based on the separate universe conjecture}

Cosmological $N$-body simulations integrate Newtonian dynamics with a changing GR metric that is calculated from averaged quantities. There is a choice in how the averaging is done:

{\em Standard approach:} Traditional cosmological $N$-body simulations use the Friedmann equations with the average density (calculated as the total mass of particles divided by the volume) to determine the overall expansion rate at each time step. Implicit in this approach is the calculation of the average density, since the total mass of particles is constant, and so is the average comoving density.

{\em Average Expansion Rate Approximation (AvERA) approach:} Using the separate universe conjecture and neglecting anisotropies around spherically symmetric regions, we calculate the local expansion rate on a grid from the Friedmann equations using the local density, and then perform spatial averaging to calculate the overall expansion rate. The algorithm exchanges the order of averaging and calculating the expansion rate and, due to the non-linearity of the equations, the two operations do not commute, see Fig~\ref{fig:scheme1}.

\begin{figure}
\begin{center}

\begin{equation}
\sbox0{$\begin{array}{ll}
        \Omega_{m,1}\\
        \Omega_{m,2}\\
        \dots \\
        \Omega_{m,N}\\
\end{array}$}
\mathopen{\resizebox{1.2\width}{\ht0}{$\Bigg\langle$}}
\usebox{0}
\mathclose{\resizebox{1.2\width}{\ht0}{$\Bigg\rangle$}} 
\Rightarrow \boxed{\mathrm {Friedmann~eq.}} \Rightarrow
\sbox1{$\begin{array}{ll}
                  \Delta V \\
\end{array}$}
\usebox{1}
\Rightarrow a(t+\delta t)
\label{eq:averaging1}
\end{equation}

\begin{equation}
\sbox0{$\begin{array}{ll}
                  \Omega_{m,1}\\
 		\Omega_{m,2}\\
		\dots \\
		 \Omega_{m,N}\\
\end{array}$}
\usebox{0}
\Rightarrow \boxed{\mathrm {Friedmann~eq.}} \Rightarrow
\sbox1{$\begin{array}{ll}
                 \Delta V_1\\
 		\Delta V_2\\
		\dots \\
		\Delta V_N\\
\end{array}$}
\mathopen{\resizebox{1.2\width}{\ht0}{$\Bigg\langle$}}
\usebox{1}
\mathclose{\resizebox{1.2\width}{\ht0}{$\Bigg\rangle$}} 
\Rightarrow a(t+\delta t)
\label{eq:averaging2}
\end{equation}

\end{center}
\caption{Top: Standard cosmological N-body simulations evolve the Friedmann equations using the average density. Since the total mass is constant the scale factor increment is independent of density fluctuations. Bottom: We calculate the expansion rate of local mini-universes and average the volume increment spatially to get the global scale factor increment.}
\label{fig:scheme1}
\end{figure}

Our collisionless cosmological $N$-body simulation code \citep{Racz2016,Racz2016b} applies the Delaunay Tessellation Field Estimation (DTFE) method \citep{DFTE_2000A&A...363L..29S} to estimate the local density $\rho_\mathcal{D}$ from the discrete particles. The output of DTFE is the density field on a regular grid of small cubes with equal volume $\mathcal{D}$. The code can compute the average scale factor increment of an inhomogeneous universe as described above, but it can also reproduce the standard $\Lambda$CDM simulation results obtained with Gadget2 \citep{Gadget2_2005MNRAS.364.1105S} when executed with the same initial conditions.

We estimate the expansion rate from the local average density
\begin{equation}
	\Omega_{m,\mathcal{D}} = \frac{\rho_{\mathcal{D}}}{\rho_{c,0}},
\label{eq:Local_Omega_m}
\end{equation}
using the matter-only Friedmann equations
\begin{equation}
	H_\mathcal{D} = \frac{\dot{a}_\mathcal{D}}{a_\mathcal{D}} = H_0\sqrt{\Omega_{m,\mathcal{D}}a_\mathcal{D}^{-3}+(1-\Omega_{m,\mathcal{D}})a_\mathcal{D}^{-2}}.
\label{eq:local_Friedamnn_eq}
\end{equation}
Note that Eqs.~\ref{eq:Local_Omega_m}~and~\ref{eq:local_Friedamnn_eq} are identical to the Newtonian spherical collapse equations which provide a surprisingly accurate description of the full dynamics \citep{2016MNRAS.455L..11N}, and form the basis of other successful approximations, such as halo models and the Press--Schecter formalism. 

The volumetric expansion of mini-universes is the cube of the linear expansion, assuming statistical isotropy. Ignoring the boundary conditions and the local environment of touching Lagrangian regions, one can average the volume increment of the independent domains to get the total volume increment of the simulation cube, i.e. the global increment of a homogeneous, effective scale factor, c.f. Eq.~\ref{eq:averaging2}. This is equivalent to neglecting correlations between regions and non-sphericity caused by tidal forces, not unlike in the case of halo models. The statistical approach means that we can avoid stitching together regions of space-time. We use a global simulation time step size and, while the corresponding infinitesimal changes of local redshift may vary from region to region,
the expansion rate is averaged in every simulation step, hence distances and velocities are rescaled homogeneously using the effective scale factor. As a result, similarly to standard N-body simulations, time is kept homogeneous and in one-to-one correspondence with redshift.

We ran simulations with up to $1.08\cdot10^6$ particles of mass $M=1.19\cdot10^{11} M_{\odot}$  in a volume of $147.62^3$~Mpc$^3$. The initial redshift was set to $z=9$ for both the standard $\Lambda$CDM and the AvERA simulations. At this redshift, backreaction and the effect of $\Lambda$ are both expected to be negligible. Since we focus on the expansion rate, Zel'dovich transients from the late start are insignificant. Initial conditions were calculated using LPTic \citep{LPTic_2006MNRAS.373..369C} with a fluctuation amplitude of $\sigma_8=0.8159$ which is defined assuming the $\Lambda$CDM growth function. The initial expansion rate was set to match the current value of $H_0=67.74~\Hunit$ \citep{Planck2015} for the $\Lambda$CDM model, yielding $H_{z=9.0}=1191.9~\Hunit$. Except for the value of $\Lambda$, AvERA simulations were run with parameters derived from the latest \textit{Planck} CMB observations.

\begin{figure}
\begin{center}
\includegraphics[width=\columnwidth]{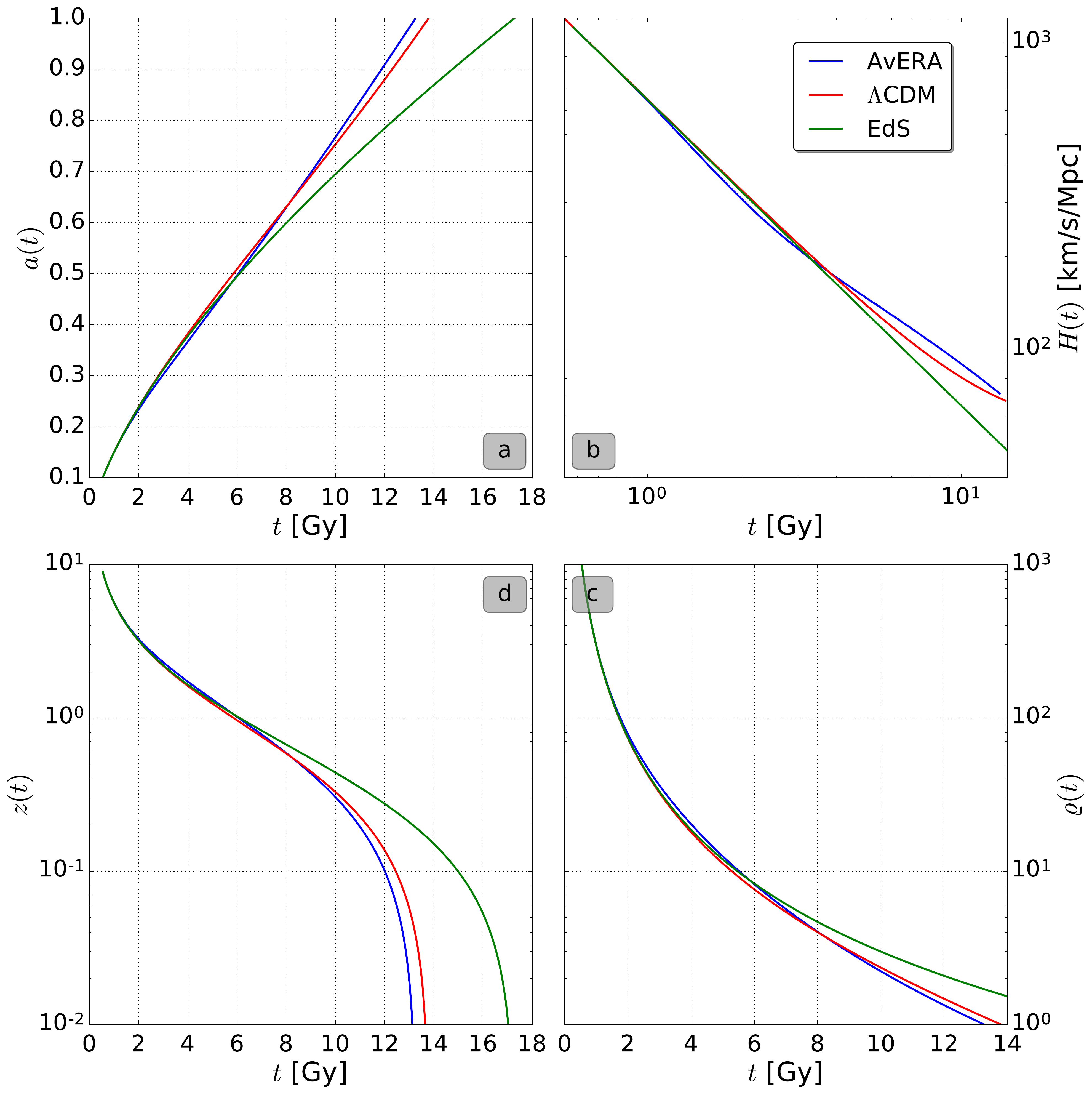}
\end{center}
\vspace*{-0.2cm}
\caption{The expansion history of the universe. Clockwise from the upper left, we plot the scale factor, the Hubble parameter, the matter density and the redshift as functions of the simulation time $t$, i.e. the age of the universe. See the text for a discussion.
}
\label{fig.at_1200}
\end{figure}

As a consistency test, the initial conditions exactly reproduce the $\Lambda$CDM expansion history when inhomogeneities are not accounted for and $\Lambda$ is non-zero. Similarly, with $\Lambda=0$ and homogeneous expansion, the initial conditions reproduce the expansion history of a flat, matter only ($\Omega_m=1$, $\Omega_{\Lambda}=0$) FLRW model with $H_0=37.69~\Hunit$. Fig.~\ref{fig.at_1200} summarizes the main results of our paper, where the expansion history $a(t)$, the Hubble parameter $H(t)$, the redshift $z(t)$ and the average density $\rho(t)$ are plotted for the AvERA model (blue), $\Lambda$CDM (red) and EdS (green) with the same initial conditions at $z=9$. The evolution of the parameters from AvERA mimic $\Lambda$CDM remarkably well, while the EdS model deviates more and more at later epochs. We emphasize that, despite the overall similarity, there are small numerical differences between the former two models which can be tested in future high precision observations.

As it was mentioned before, in AvERA simulations the expansion history and the resulting present day Hubble parameter depend on the particle mass, which corresponds to the coarse graining scale. To explore the effects of the coarse graining scale, we executed simulations with different particle masses between $1.17 \cdot 10^{11} - 3 \cdot 10^{12} M_{\odot}$. The resulting $z=0$ Hubble parameters, as a function of particle mass, are summarised in Tab.~\ref{tab:parameters}. The sensitivity of $H_0$ to the coarse graining scale is relatively minor: a factor of 10 change in the particle mass causes about a 10~per~cent change in the present time Hubble parameter, see Fig.~\ref{fig:H0HSTART}. 
The detailed investigation of this effect will be presented in a future paper.

\section{Comparison with observations}

Given the close similarity of the expansion history of the AvERA model with that of $\Lambda$CDM, and the fact that linear growth history is driven by the time evolution of the expansion rate, the AvERA model provides an adequate framework for the interpretation of many observations supporting the concordance model, despite the fact that the current version of the simulation is not suitable yet to compute light propagation across the curved space-time regions. Luckily, luminosity distance at low redshift (but beyond the statistical scale of homogeneity) is primarily determined by the expansion history and is only slightly sensitive to curvature. In what follows, we do not attempt to fit any data, we simply plot our fiducial model with different coarse graining scales against select key observations.

\begin{figure}
\begin{center}
\includegraphics[width=0.6\columnwidth]{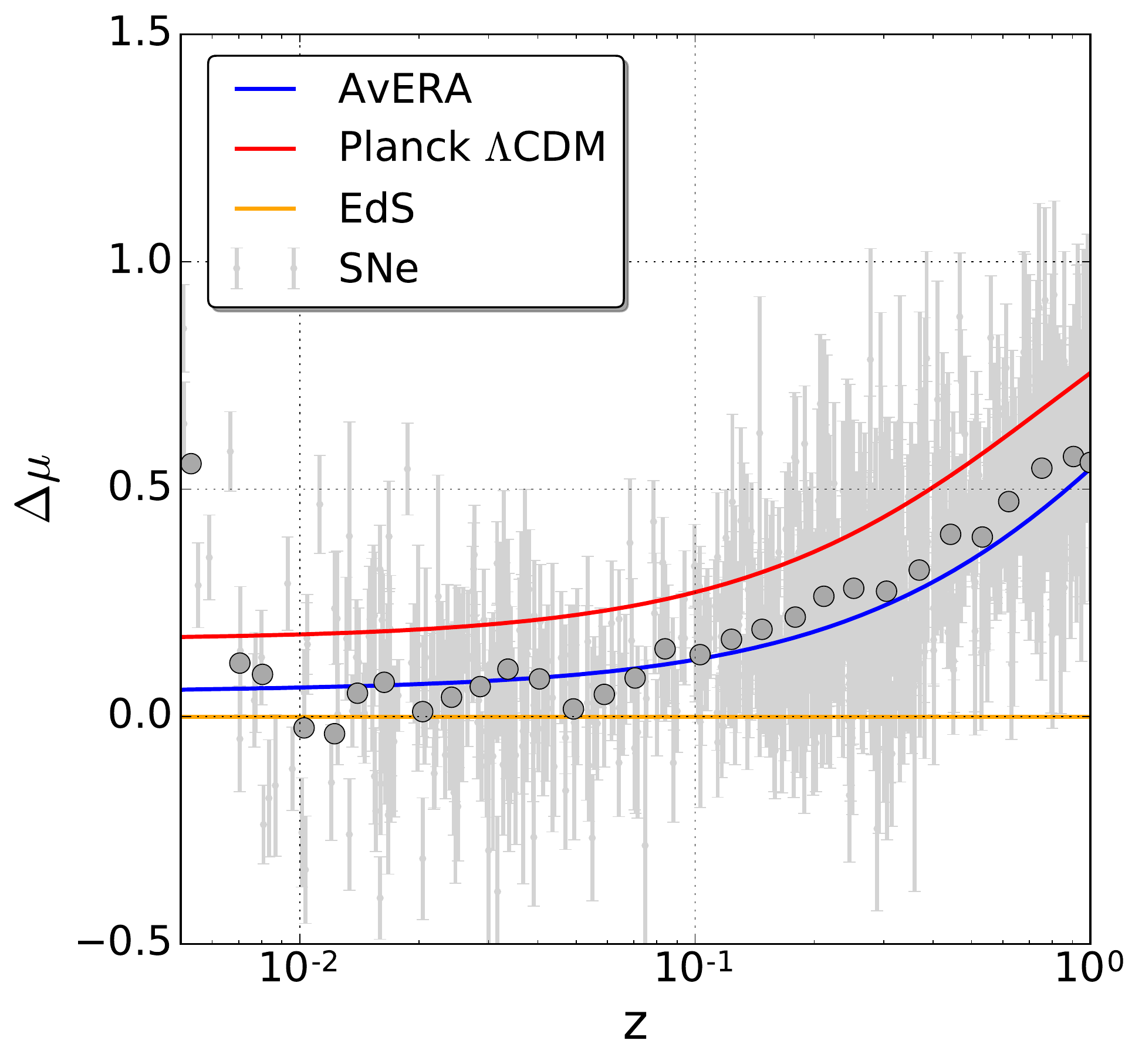}
\vspace*{-0.25cm}
\end{center}
\caption{The relative distance modulus $\Delta \mu = DM - DM_{EdS}$ as a function of redshift. The green line corresponds to the reference $\Omega_{m}=1$ flat Universe, while the red curve shows the standard $\Lambda$CDM model with the concordance cosmological parameter set. The observed values from the SuperCal supernova compilation \citep{Scolnic2015ApJ...815..117S, Scolnic2016ApJ...822L..35S}, calibrated using Cepheid distances by Riess et al. \citep{Riess_hubbleParameterTension_2016}, are shown with gray errorbars, and their binned values with darker dots. The result of our second  $\Omega_{m}=1$ AvERA simulation with a particle mass of $2.03 \cdot10^{11} M_{\odot}$ (see also Fig.~\ref{fig:H0HSTART} and Tab.~\ref{tab:parameters}) is shown in blue. It differs from the homogeneous EdS solution, and fits better the supernova data than the concordance $\Lambda$CDM model.}
\label{fig:delta_dm1191}
\end{figure}

\begin{figure}
\begin{center}
\includegraphics[width=\columnwidth]{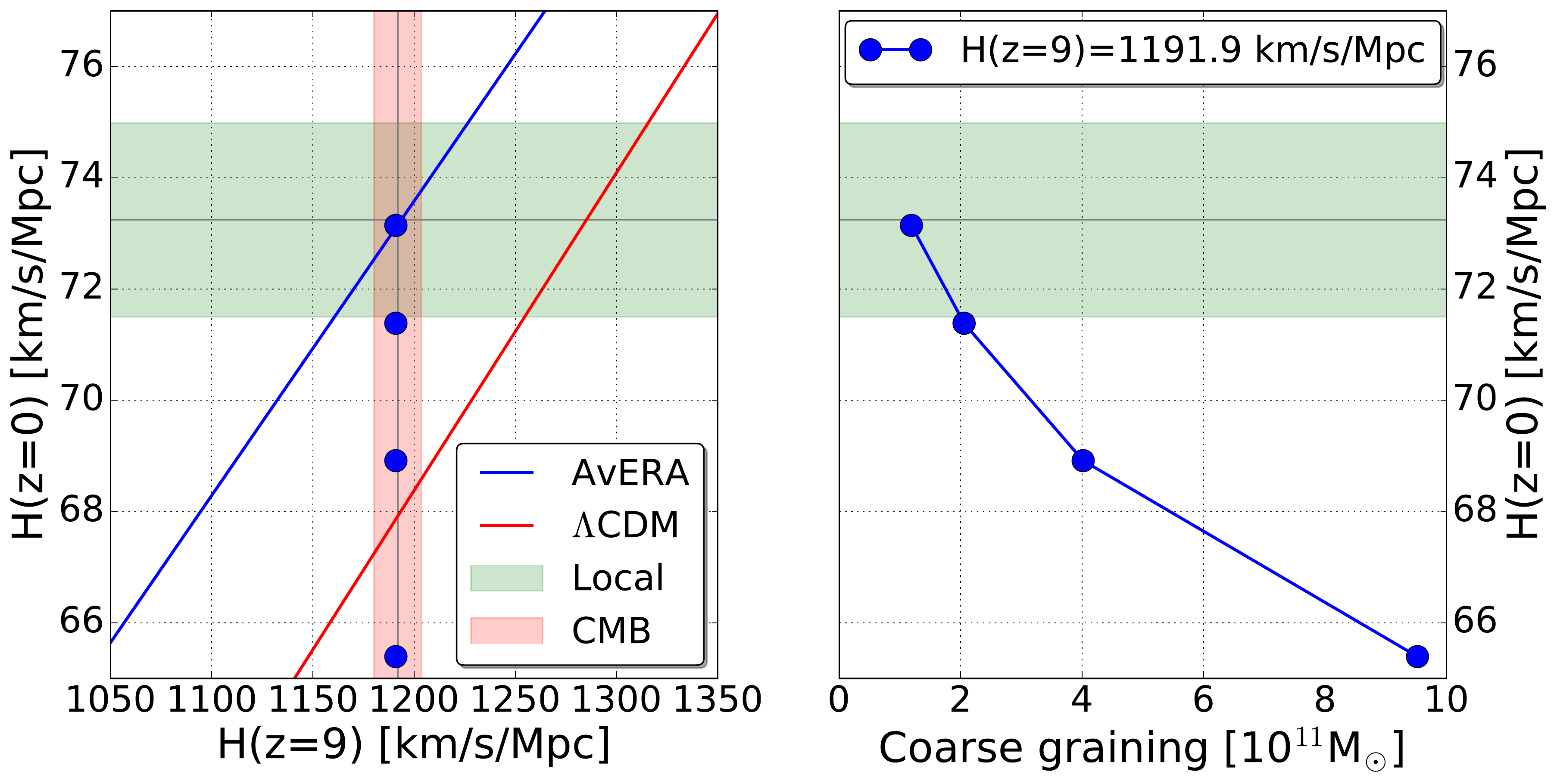}
\vspace*{-0.5cm}
\end{center}
\caption{ Left: The relation between the Hubble parameter values for the local (vertical axis) and the distant (horizontal axis) Universe. The horizontal stripe corresponds to the 1$\sigma$ range allowed by the most recent local calibration \citep{Riess_hubbleParameterTension_2016}, while the vertical one is the 1$\sigma$ range calculated for $z=9$ from the latest CMB measurements \citep{Planck2015}. N-body simulations started at different $H(z=9)$ values for the $\Lambda$CDM model (red line) cannot satisfy both criteria (intersection of stripes), while at a reasonable coarse graining scale our simulation (blue) fits both observations. Right: The effect of coarse graining on $H_0$ at a given initial Hubble parameter $H(z=9)$. }
\label{fig:H0HSTART}
\end{figure}

One of the first and strongest observational proofs of accelerating cosmic expansion came from type~Ia supernova distance modulus measurements \citep{Riess_1998AJ....116.1009R, Perlmutter_1999ApJ...517..565P, Scolnic2015ApJ...815..117S}. Fig.~\ref{fig:delta_dm1191} shows the distance moduli from the observations overplotted with curves from the EdS, \textit{Planck} $\Lambda$CDM, and our model. We used the SuperCal compilation \citep{Scolnic2015ApJ...815..117S, Scolnic2016ApJ...822L..35S} of supernova observations, with magnitudes corrected to the fiducial color and luminosity, and set the zero point of the absolute magnitude scale to match the Cepheid-distance-based absolute magnitudes as determined by Riess et al. \citep{Riess_hubbleParameterTension_2016}. Both the \textit{Planck} $\Lambda$CDM and our AvERA model follow the observed deviation from EdS. If we choose the coarse graining scale such that the local Hubble constant is reproduced (see Fig.~\ref{fig:H0HSTART}), the AvERA model is favoured: $\chi^2_\textnormal{AV}=1347.8$ vs. $\chi^2_{\Lambda\textnormal{CDM}}=2485.7$, see Fig~\ref{fig:delta_dm1191}.

The tension between the locally measured \citep{Riess_hubbleParameterTension_2016} value of the Hubble constant $H_0=73.24\pm1.74~\Hunit$ and the estimate from \textit{Planck} data \citep{Planck2015} $H_0=67.27\pm0.66~\Hunit$ is worthy of special attention, given that the significance of the difference is over $3\sigma$ within the $\Lambda$CDM paradigm. According to Fig.~\ref{fig:H0HSTART} which displays the high and low $z$ constraints, our model can naturally incorporate both. In particular, our two highest resolution simulations that are consistent with the \textit{Planck} constraints yield the values of  $71.38~\Hunit$ and $73.14~\Hunit$ for the Hubble constant, respectively. At the same time, since we calculate the Hubble parameter as a spatial average over the universe, we cannot account for the effect of inhomogeneities on $H_0$ on scales smaller than the statistical homogeneity scale which could explain the tension between local and CMB observations.

\section{Conclusions and Discussion}

We have presented a modified $N$-body simulation where we estimated the global expansion rate by averaging local expansion rates of mini-universes based on the separate universe approximation. While we do not attempt to connect space-time regions or compute light propagation across curved regions, our approach is equivalent to a non-perturbative statistical backreaction calculation. Our approximation neglects tidal forces due to anisotropies, and has an ambiguity associated with the optimal coarse graining scale. For a large enough scale, the effect is negligible, while on small scales anisotropies break the underlying assumptions of the approximation. Since virialised objects detach from the expansion, we expect that the optimal coarse graining scale, treated as a nuisance parameter, is related to the size of the typical virialised regions.

\begin{table}
\begin{tabular}{r | c | c}
$N$ & $M \left[ 10^{11} M_\odot \right]$ & $H_0~\left[ \Hunit \right]$ \\
\hline
135,000 & 9.40 & 65.4 \\
320,000 & 3.96 & 68.9 \\
625,000 & 2.03 & 71.4 \\
1,080,000 & 1.17 & 73.1 \\
\end{tabular}
\caption{Summary of simulation input parameters and the resulting values of $H_0$. In all cases the linear size of the simulation box was $L=147.62~\unit{Mpc}$ and the early epoch value of the Hubble parameter was set to $H_{z=9} = 1191.9~\Hunit$, complying with \textit{Planck} $\Lambda$CDM best-fit parameters.}
\label{tab:parameters}
\end{table}

Our modified $\Omega_m=1$ simulation mimics the $\Lambda$CDM expansion history remarkably well. Since growth history is also driven by the expansion history, we expect that our simulations are consistent with luminosity distance and Hubble parameter observations constraining dark energy. Present-day supernova observations are well fit by our model, moreover, our model naturally resolves the tension between local and CMB Hubble constant measurements. Detailed fits to observations, and forecasting for future surveys such as  Euclid, WFIRST, HSC, etc. is left for future work, but it is clear already from Fig.~\ref{fig.at_1200} that if our model is sufficiently different from the standard $w=-1$ vacuum energy model, upcoming surveys will be able to confirm or rule it out. We also note that some of our results are numerically very similar to the analytically derived timescape scenario presented in \cite{Wiltshire2007b, Wiltshire2009}, in addition to sharing the separate universe approximation. Further investigation is yet to be done to compare the two approaches in detail.

\begin{figure}
\begin{center}
\includegraphics[width=\columnwidth]{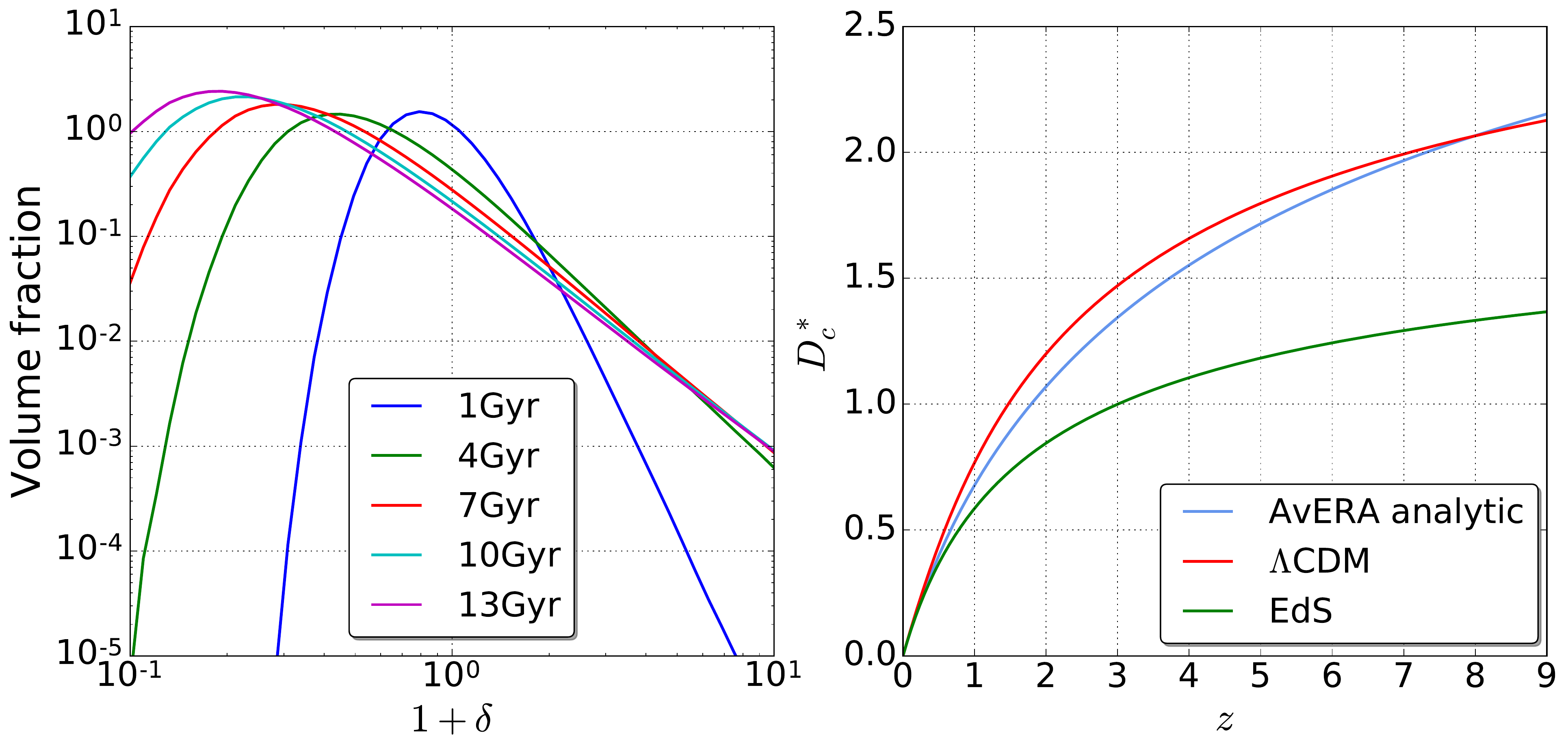}
\vspace*{-0.5cm}
\end{center}
\caption{Left: Evolution of the distribution of $1+\delta$ during the AvERA N-body simulation. Right: The normalized line of sight comoving distance $D_c^*= \int_0^z H_0/(H(z^\prime))dz^\prime$ at each time step has been calculated from a matter-only FLRW model with $\Omega_m$ corresponding to the actual peak $1+\delta$ (blue). The curve deviates from the  $\Omega_m=1$ model (green) and closely follows the $\Lambda$CDM model (red).}
\label{fig:OmegaEvol}
\end{figure}

To investigate the validity of the AvERA simulation, we performed follow-up analytic calculations. If inhomogeneities mainly affect structure growth via the expansion history of the universe, the (near) lognormal approximation found in $\Lambda$CDM simulations should approximate well the density distribution of the mini-universes. For instance, to calculate the longitudinal comoving scale, we calculate the average
\begin{equation}
\frac{1}{H_\textnormal{eff}(z)} = 
\avg{\frac{1}{H(z)}} = \int_{-1}^{\infty} P(1+\delta) \frac{1}{H_{\mathcal{D}}(1+\delta, z)} \,\textnormal{d}\delta,
\end{equation}
where $P(1+\delta)$ is a lognormal distribution, and $H_{\mathcal{D}}(1+\delta, z)$ is calculated from Equation~\ref{eq:local_Friedamnn_eq} with $\Omega = 1+\delta$. The variance of the lognormal distribution is estimated from $\sigma_A^2 = 0.73\log (1+\sigma^2_{lin}/0.73)$ \citep{ReppSzapudi2017}, and $\avg{\log(1+\delta)} = - 0.67 \log(1+\sigma^2_\textnormal{lin}/(2\times 0.67))$ \citep{ReppSzapudi2017}, where $\sigma_\textnormal{lin}$ is the linear variance of dark matter fluctuations on the coarse graining scale. The lognormal PDF is a good approximation and the above fits are accurate for concordance cosmologies. The result of such a calculation is shown on the right panel of Figure~\ref{fig:OmegaEvol}. Details will be presented elsewhere (Szapudi et.al. 2017 in prep.).

The theoretical calculation is insensitive to the details of the PDF, i.e. departures from lognormality. We can calculate the effective expansion rate fairly accurately by replacing $\Omega_\textnormal{M}$ with its most likely value, i.e. $1+\delta$ at the peak of the density PDF on the left panel of Figure~\ref{fig:OmegaEvol}. This corresponds to approximating the lognormal PDF with a Dirac-$\delta$ function centred on the peak of the distribution. Thus the physical meaning of these calculations is simple: according to our approximation, it is not the average but the typical energy density that governs the expansion rate of the Universe. At high redshifts, where the distribution is fairly symmetric, the typical value of $\delta$ (mode of the PDF) is close to the average and the Universe evolves without backreaction. At late times skewness increases, the volume of the Universe is dominated by voids, and the typical value of $\delta$ is negative, thus effectively $\Omega_\textnormal{M} < 1$. High density regions, where metric perturbations are perhaps the largest, are inconsequential to this effect: what matters is the non-Gaussianity of the density distribution, in particular, the large volume fraction of low density regions, as advertised earlier.

The statistical approach we use is spatial (volume) averaging. While averaging is ambiguous in curved space-times \citep{Wiltshire2014}, note that all astrophysical quantities, most notably the power spectrum that is used to calculate all cosmological parameters, are estimated through analogous statistical procedures. We neglect local anisotropies, and we assume that those effects average out over time. Nevertheless, one could generalize our code to include tidal forces using elliptical collapse equations for the mini-universes we consider. This refinement of our calculations would quantify to first order the effect from tidal forces, but is left for future work. Our simple model with a reasonable coarse graining scale yields results that are consistent with the observations, and the model also has a simple physical interpretation. Further studies are needed both on the theoretical front and on fitting cosmological parameters, nevertheless, our approach is not only a viable alternative to dark energy models, but appears to be flexible enough to resolve some tensions in a natural way.

\section{Acknowledgements}
This work was supported by NKFI NN 114560. IS acknowledges NASA grants NNX12AF83G and NNX10AD53G for support. RB was
supported through the New National Excellence Program of the Ministry of Human Capacities, Hungary. The authors thank Daniel Scolnic for providing the filtered supernova sample and feedback concerning its details. We also thank Alex Szalay and Mark Neyrinck for stimulating discussions and comments.
 
\bibliographystyle{mn2e}
\bibliography{inhomExpans,inhomExpansNotes}

\begin{thebibliography}{32}
\expandafter\ifx\csname natexlab\endcsname\relax\def\natexlab#1{#1}\fi

\bibitem[{{Bernardeau}(1994)}]{Bernardeau1994}
{Bernardeau} F., 1994, \apj, 427, 51

\bibitem[{{Buchert}(2000)}]{Buchert2000}
{Buchert} T., 2000, General Relativity and Gravitation, 32, 105

\bibitem[{{Buchert}(2001)}]{Buchert2001}
{Buchert} T., 2001, General Relativity and Gravitation, 33, 1381

\bibitem[{{Buchert} {et~al}\mbox{.}(2015){Buchert}, {Carfora}, {Ellis}, {Kolb},
  {MacCallum}, {Ostrowski}, {R{\"a}s{\"a}nen}, {Roukema}, {Andersson}, {Coley},
  \& {Wiltshire}}]{2015CQGra..32u5021B}
{Buchert} T. {et~al.}, 2015, Classical and Quantum Gravity, 32, 215021

\bibitem[{{Buchert} {et~al}\mbox{.}(2016){Buchert}, {Coley}, {Kleinert},
  {Roukema}, \& {Wiltshire}}]{Buchert_2016IJMPD..2530007B}
{Buchert} T., {Coley} A.~A., {Kleinert} H., {Roukema} B.~F., {Wiltshire} D.~L.,
  2016, International Journal of Modern Physics D, 25, 1630007

\bibitem[{{Buchert} \& {R{\"a}s{\"a}nen}(2012)}]{2012ARNPS..62...57B}
{Buchert} T., {R{\"a}s{\"a}nen} S., 2012, Annual Review of Nuclear and Particle
  Science, 62, 57

\bibitem[{{Crocce}, {Pueblas} \& {Scoccimarro}(2006){Crocce}, {Pueblas}, \&
  {Scoccimarro}}]{LPTic_2006MNRAS.373..369C}
{Crocce} M., {Pueblas} S., {Scoccimarro} R., 2006, MNRAS, 373, 369

\bibitem[{{Dai}, {Pajer} \& {Schmidt}(2015){Dai}, {Pajer}, \&
  {Schmidt}}]{Dai2015}
{Dai} L., {Pajer} E., {Schmidt} F., 2015, \jcap, 10, 059

\bibitem[{Giblin, Mertens \& Starkman(2016)Giblin, Mertens, \&
  Starkman}]{PhysRevLett.116.251301}
Giblin J.~T., Mertens J.~B., Starkman G.~D., 2016, Phys. Rev. Lett., 116,
  251301

\bibitem[{{Green} \& {Wald}(2014)}]{2014CQGra..31w4003G}
{Green} S.~R., {Wald} R.~M., 2014, Classical and Quantum Gravity, 31, 234003

\bibitem[{{Green} \& {Wald}(2016)}]{2016CQGra..33l5027G}
{Green} S.~R., {Wald} R.~M., 2016, Classical and Quantum Gravity, 33, 125027

\bibitem[{{Kolb} {et~al}\mbox{.}(2005){Kolb}, {Matarrese}, {Notari}, \&
  {Riotto}}]{Kolb2005}
{Kolb} E.~W., {Matarrese} S., {Notari} A., {Riotto} A., 2005, \prd, 71, 023524

\bibitem[{{Mohayaee} {et~al}\mbox{.}(2006){Mohayaee}, {Mathis}, {Colombi}, \&
  {Silk}}]{Mohayaee2006}
{Mohayaee} R., {Mathis} H., {Colombi} S., {Silk} J., 2006, \mnras, 365, 939

\bibitem[{{Neyrinck}(2016)}]{2016MNRAS.455L..11N}
{Neyrinck} M.~C., 2016, \mnras, 455, L11

\bibitem[{{Perlmutter} {et~al}\mbox{.}(1999){Perlmutter}, {Aldering},
  {Goldhaber}, {Knop}, {Nugent}, {Castro}, {Deustua}, {Fabbro}, {Goobar},
  {Groom}, {Hook}, {Kim}, {Kim}, {Lee}, {Nunes}, {Pain}, {Pennypacker},
  {Quimby}, {Lidman}, {Ellis}, {Irwin}, {McMahon}, {Ruiz-Lapuente}, {Walton},
  {Schaefer}, {Boyle}, {Filippenko}, {Matheson}, {Fruchter}, {Panagia},
  {Newberg}, {Couch}, \& {Project}}]{Perlmutter_1999ApJ...517..565P}
{Perlmutter} S. {et~al.}, 1999, \apj, 517, 565

\bibitem[{{Planck Collaboration}(2016)}]{Planck2015}
{Planck Collaboration}, 2016, \aap, 594, A13

\bibitem[{{R\'acz, G. et al.}(2016)}]{Racz2016b}
{R\'acz, G. et al.}, 2016, source code available at
  {\url{http://www.vo.elte.hu/papers/2016/nbody/}}

\bibitem[{{R\'acz, G. et al.}(2017)}]{Racz2016}
{R\'acz, G. et al.}, 2017, in prep.

\bibitem[{{R{\"a}s{\"a}nen}(2010)}]{Rasanen2010}
{R{\"a}s{\"a}nen} S., 2010, \prd, 81, 103512

\bibitem[{{Repp} \& {Szapudi}(2017)}]{ReppSzapudi2017}
{Repp} A., {Szapudi} I., 2017, \mnras, 464, L21

\bibitem[{{Riess} {et~al}\mbox{.}(1998){Riess}, {Filippenko}, {Challis},
  {Clocchiatti}, {Diercks}, {Garnavich}, {Gilliland}, {Hogan}, {Jha},
  {Kirshner}, {Leibundgut}, {Phillips}, {Reiss}, {Schmidt}, {Schommer},
  {Smith}, {Spyromilio}, {Stubbs}, {Suntzeff}, \&
  {Tonry}}]{Riess_1998AJ....116.1009R}
{Riess} A.~G. {et~al.}, 1998, \aj, 116, 1009

\bibitem[{{Riess} {et~al}\mbox{.}(2016){Riess}, {Macri}, {Hoffmann}, {Scolnic},
  {Casertano}, {Filippenko}, {Tucker}, {Reid}, {Jones}, {Silverman},
  {Chornock}, {Challis}, {Yuan}, {Brown}, \&
  {Foley}}]{Riess_hubbleParameterTension_2016}
{Riess} A.~G. {et~al.}, 2016, \apj, 826, 56

\bibitem[{{Schaap} \& {van de Weygaert}(2000)}]{DFTE_2000A&A...363L..29S}
{Schaap} W.~E., {van de Weygaert} R., 2000, A\&A, 363, L29

\bibitem[{{Scolnic} {et~al}\mbox{.}(2015){Scolnic}, {Casertano}, {Riess},
  {Rest}, {Schlafly}, {Foley}, {Finkbeiner}, {Tang}, {Burgett}, {Chambers},
  {Draper}, {Flewelling}, {Hodapp}, {Huber}, {Kaiser}, {Kudritzki}, {Magnier},
  {Metcalfe}, \& {Stubbs}}]{Scolnic2015ApJ...815..117S}
{Scolnic} D. {et~al.}, 2015, \apj, 815, 117

\bibitem[{{Scolnic} \& {Kessler}(2016)}]{Scolnic2016ApJ...822L..35S}
{Scolnic} D., {Kessler} R., 2016, \apjl, 822, L35

\bibitem[{{Springel}(2005)}]{Gadget2_2005MNRAS.364.1105S}
{Springel} V., 2005, \mnras, 364, 1105

\bibitem[{Weinberg(2008)}]{Weinberg2008}
Weinberg S., 2008, Cosmology. Oxford University Press, Oxford, UK

\bibitem[{{Wiltshire}(2007{\natexlab{a}})}]{Wiltshire2007a}
{Wiltshire} D.~L., 2007{\natexlab{a}}, New Journal of Physics, 9, 377

\bibitem[{{Wiltshire}(2007{\natexlab{b}})}]{Wiltshire2007b}
{Wiltshire} D.~L., 2007{\natexlab{b}}, Physical Review Letters, 99, 251101

\bibitem[{{Wiltshire}(2009)}]{Wiltshire2009}
{Wiltshire} D.~L., 2009, \prd, 80, 123512

\bibitem[{Wiltshire(2011)}]{WiltshireCQG}
Wiltshire D.~L., 2011, Classical and Quantum Gravity, 28, 164006

\bibitem[{{Wiltshire}(2014)}]{Wiltshire2014}
{Wiltshire} D.~L., 2014, in Proceedings of the 15th Brazilian School on
  Cosmology and Gravitation, Perez~Bergliaffa S.E.;~Novello M., ed., Cambridge
  Scientific Publishers, Cambridge, UK, pp. 203--244

\end{thebibliography}

\end{document}